\documentclass[preprintnumbers,amsmath,amssymb,twocolumn]{revtex4}

\usepackage{amsmath}
\usepackage{amsfonts}
\usepackage{amssymb}
\usepackage{graphicx}

\begin{document}

\title{Single molecule pulling with large time steps}
\author{Harald Oberhofer}
\author{Christoph Dellago}
\affiliation{Faculty of Physics and Center for Computational Materials Science, University of Vienna, Boltzmanngasse 5, 1090 Vienna, Austria}

\author{Stefan Boresch}
\affiliation{Faculty of Chemistry, Institute for Computational Biological Chemistry, University of Vienna, W\"ahringer Strasse 17, 1090 Vienna, Austria}
\date{\today}


\begin{abstract}
Recently, we presented a generalisation of the Jarzynski
non-equilibrium work theorem for phase space mappings. The formalism
shows that one can determine free energy differences from approximate
trajectories obtained from molecular dynamics simulations in which
very large timesteps are used. In this work we test the method by
simulating the force induced unfolding of a deca-alanine helix in
vacuum. The excellent agreement between results obtained with a small,
conservative time step of 0.5~fs and results obtained with a time step
of 3.2~fs (i.e., close to the stability limit) indicates that the
large time step approach is practical for such complex biomolecules.
We further adapt the method of Hummer and Szabo for the simulation of
single-molecule force spectroscopy experiments to the large time step
method. While trajectories generated with large steps are approximate
and may be unphysical --- in the simulations presented here we observe
a violation of the equipartition theorem --- the computed free
energies are exact in principle. In terms of efficiency, the optimum
time step for the unfolding simulations lies in the range 1--3~fs.
\end{abstract}

\maketitle


\section{Introduction}

Modern experimental techniques such as atomic force microscopy or
optical tweezers offer the means to study the mechanical properties of
single molecules such as proteins or nucleic acids providing striking
insights into their structure, energetics and dynamics
\cite{Rief1997,Marszalek1999,Liphardt2002}. In such experiments,
finely tuned forces are used to distort individual molecules and to
track their response with high resolution. Although mechanical single
molecule experiments can yield a wealth of useful information, their
thermodynamic analysis in terms of binding constants or unfolding free
energies is not straightforward. This complication originates from the
fact that typically the perturbation acting on the system is time
dependent, driving the system away from equilibrium. In such a
non-equilibrium situation the free energy, or reversible work, cannot
be simply calculated by integration of the average force along the
pulling path. A solution is offered by Jarzynski's theorem which
relates equilibrium free energies to the statistics of work carried
out during non-equilibrium transformations
\cite{Jarzynski1,Jarzynski2}.

According to the Jarzynski equality, the free energy difference
$\Delta F$ between two states corresponding to different values of an
externally controlled parameter $\lambda$ can be obtained from an
exponential average of the work $W$ performed on the system by
switching the parameter from its initial to its final value:
	\begin{equation}
		\label{eq:jarz}
		e^{-\beta \Delta F} = \left\langle e^{ -\beta W} \right\rangle.
	\end{equation}
The average, denoted by the angular brackets, extends over all
realisations of the switching process starting from a system initially
in equilibrium with a heat bath at temperature $T=1/k_{\rm B}T$. In an
experiment in which a single molecule, say a DNA fragment, is pulled
by an optical trap, the external parameter $\lambda$ corresponds to
the trap position and $W$ is the work performed on the system by the
moving trap. Since the force exerted on the molecule by the optical
trap can be measured for every trap position, the work $W$ can be
calculated from the experimental data and the average from
Equ.~(\ref{eq:jarz}) can be determined by repeating the pulling
experiment many times.

While such an analysis of the experimental data is possible, one is
rarely interested in the free energy of the entire system (molecule
plus trap) as a function of the trap position. Rather, the interest
usually lies in the equilibrium free energy of the untrapped molecule
as a function of some conformational degree of freedom such as the
end-to-end distance. A way to extract that information from
non-equilibrium work data has been recently proposed by Hummer and
Szabo \cite{HummerSzabo1,HummerSzabo2}. This method is based on the
insight that the equilibrium distribution from which the
transformation is initiated can be reconstructed from the
non-equilibrium distribution generated in the process. This
reconstruction, in which the bias of the trap is removed, requires
only knowledge of the work carried out in the transformation as well
as the time-dependent trap potential, both quantities that are
accessible experimentally.

Information extracted from single molecule experiments can be
complemented with corresponding non-equilibrium computer simulations
which provide detailed atomistic pictures of the molecule's response
to the mechanical perturbation \cite{Paci,Schulten,Grubmueller}. As is
the case for the experiments, the calculation of
equilibrium free energies from such simulations using Jarzynski's
method or Hummer and Szabo's refinement becomes increasingly difficult
if the external perturbation strongly removes the system from
equilibrium. In this regime, the exponential average from
Equ.~(\ref{eq:jarz}) is dominated by a few rare trajectories leading
to large statistical uncertainties in the free energy estimate
\cite{Jarzynski2006}. This difficulty may be overcome in computer
simulations by generating trajectories yielding rare but important
work values with enhanced likelihood, and several techniques to do
that have been devised
\cite{Oberhofer1,Sun1,Ytreberg1,WuKofke1}. Using these biased sampling
methods, the number of non-equilibrium trajectories required to
calculate free energies with a given accuracy can be dramatically
lowered.

A different type of efficiency enhancement of such
non-equilibrium fast-switching simulations can be achieved by reducing
the computational cost required to generate the individual
trajectories. We have recently proposed a method to do
that based on the integration of the equations of motion with
unusually large time steps \cite{Lechner1,Oberhofer2006}. Since the
number of steps required to propagate the system for a given time
decreases with increasing time step, this approach promises
computational savings. Although approximate trajectories obtained with
large time steps mimic the true dynamics of the system only crudely,
the resulting free energies are in principle exact, the
obvious limitation being the numerical stability of the integrator.
In this paper, we investigate how this large-time step formalism
performs in the simulation of single molecule pulling experiments. To
do that we first show that the Hummer-Szabo procedure can be applied
without major changes to trajectories obtained with large time
steps. We then analyse the efficiency of the method for the force
induced unfolding of deca-alanine and calculate the free energy as a
function of the end-to-end distance. These simulations indicate that
for time steps ranging from a conservative time step of 0.5~fs to a
time step of 3.2~fs, just short of the stability limit, the
computational efficiency is essentially constant. The
time steps typically used in simulations of such systems are in the
range of 1--2~fs. A detailed analysis of the trajectories obtained
with different time steps reveals clear indications of their
approximate character. In particular, we observe a sudden temperature
drop of up to 5\% during the first few steps of the pulling
trajectories, as the time step is increased. This effect is most
likely because of the so-called shadow Hamiltonian that is conserved
for symplectic integrators such as the velocity-Verlet algorithm used
in this study.
	
The remainder of the paper is organised as follows. First, in
Sec.~\ref{sec:ltintro}, we review our large timestep fast-switching
method and show how the Hummer-Szabo procedure can be applied to this
case. After that we describe the model system in
Sec.~\ref{sec:model}. Results are presented in Sec.~\ref{sec:results}
followed by conclusions in Sec.~\ref{sec:conclusions}.


\section{Large time step fast switching}
\label{sec:ltintro}

Any deterministic time evolution in phase space can be viewed as a mapping that takes the initial point of a trajectory into its final point. Expanding on this perspective, we have recently derived a generalised version of the Jarzynski identity that permits to calculate exact free energies from approximate large time step trajectories both for deterministic as well as stochastic dynamics \cite{Lechner1}. The basis for this approach is provided by the following identity that can be shown to hold for a system with parameter dependent Hamiltonian $\mathcal{H}(x, \lambda)$ and a general invertible and differentiable mapping $\phi(x)$ acting on the phase space point $x$:
\begin{equation}
\label{eq:ltmain}
e^{-\beta \Delta F} = \left\langle e^{-\beta W_\phi(x)}\right\rangle.
\end{equation}
Here, the angular brackets denote an average over the equilibrium (canonical) distribution of the initial state. The work function $W_{\phi}$, defined as 
\begin{equation}
\label{eq:ltwork}
W_\phi(x)= \mathcal{H}(\phi(x),\lambda_\text{B})-\mathcal{H}(x,\lambda_\text{A})-\beta^{-1}\ln\left\vert \frac{\partial \phi(x)}{\partial x}\right\vert
\end{equation}
includes the total energy change caused by switching the parameter $\lambda$ from its initial value $\lambda_A$ to its final value $\lambda_B$. In addition, a term depending on the Jacobian $\vert\partial \phi(x)/\partial x\vert$ of the mapping contributes to the work function. If the mapping $\phi(x)$ conserves the canonical distribution, this entropic term, which takes into account the expansion or compression of phase space volume, corresponds to the heat exchange \cite{Lisi2006}. In this case, Equ.~(\ref{eq:ltwork}) is equivalent to the first law of thermodynamics. 

To apply Equ.~(\ref{eq:ltwork}) to the case of large time step trajectories, one has to consider the mapping that results from a concatenation of molecular dynamics steps. The work function then consists of the difference in total energy between the initial and final point of the trajectory and a sum of terms depending on the phase space volume change during each time step. We emphasise that in principle Equ.~(\ref{eq:ltwork}) yields exact free energy differences regardless of the size of the time step. 

The application of the large time step method becomes particularly simple if the algorithm used for the integration of the equations of motion conserves phase space volume. In this case, the Jacobian is strictly unity for each time step and the work function reduces to the energy difference accumulated along the trajectory, $W_\phi(x)= \mathcal{H}(\phi(x),\lambda_\text{B})-\mathcal{H}(x,\lambda_\text{A})$. In the present work we will only use the velocity-Verlet algorithm, which is volume-preserving \cite{FrenkelSmit1,Tuckerman92}. A detailed discussion of the large time step formalism for non volume-preserving integrators and related complications can be found in Ref.~\cite{Oberhofer2006}.

Since Jarzynski's identity holds for a time evolution resulting from the integration of the equations of motion with large time step, one expects that also the Hummer-Szabo procedure mentioned above remains valid in this case. We next show that this is indeed the case. Consider a system with Hamiltonian 
\begin{equation}
\mathcal{H}(x,\lambda_B)=\mathcal{H}_0(x)+V(q(x); \lambda),
\end{equation}
where $\mathcal{H}_0(x)$ is the molecular Hamiltonian and $V(q(x), \lambda)$ is the parameter dependent trap potential coupling to the variable $q(x)$. What one would like to calculate is the free energy $F(q)$ defined up to a constant as
\begin{equation}
e^{-\beta F(q)}=\langle \delta [q-q(x)] \rangle_0,
\end{equation}
where the brackets $\langle \cdots \rangle_0$ denote an average over the equilibrium distribution of the molecular system without the trap, i.e., over the distribution
\begin{equation}
\rho_0(x)=\frac{e^{-\beta \mathcal{H}_0(x)}}{\int dx\,  e^{-\beta \mathcal{H}_0(x)}}.
\end{equation}
The goal now is to derive an expression that permits to compute the free energy from the phase space distribution resulting from application of the mapping $\phi(x)$ where at the same time the parameter $\lambda$ is changed from its initial value $\lambda_A$ to its final value $\lambda_B$. To do so, we first perform a transformation of variables from $x$ to $y=\phi^{-1}(x)$ obtaining
\begin{equation}
e^{-\beta F(q)}=\frac{1}{Q_0}\int dy \left| \frac{\partial \phi}{\partial y}\right| e^{-\beta \mathcal{H}_0(\phi(y))} \delta [q-q(\phi(y))],
\end{equation}
where $Q_0=\int dx\,  \exp [-\beta \mathcal{H}_0(x)]$ is the partition function of the system without trap. Multiplication and division of the integrand with $\exp\{-\beta \mathcal{H}(y,\lambda_A)]-\beta V(q(y), \lambda_B)\}$ then yields
\begin{equation}
\label{equ:HuSza1}
e^{-\beta F(q)}=\frac{Q(\lambda_A)}{Q_0} e^{\beta V(q, \lambda_B)}\langle \delta [q-q(\phi(y))] e^{-\beta W_\phi}\rangle,
\end{equation}
where
\begin{equation}
Q(\lambda_A)=\int dx e^{-\beta \mathcal{H}(y,\lambda_A)}
\end{equation}
is the partition function of the entire system, including the trap at position $\lambda_A$ and the average is over initial conditions canonically distributed with respect to $\mathcal{H}(x, \lambda_A)$. Equation (\ref{equ:HuSza1}) implies that the entire free energy profile $F(q)=-k_{\rm B}T \ln \langle \delta [q-q(x)]\rangle_0$ can be calculated up to a constant by histogramming the variable $q$ at the end of the transformation $\phi(x)$ and weighing each contribution to the histogram with the work exponential $\exp(-\beta W_\phi)$. This weight essentially takes into account the different phase space probability of a particular point in the equilibrium ensemble and the ensemble generated by the mapping. Equation (\ref{equ:HuSza1}) is analogous to the central result of Hummer and Szabo (Equ.~(7) of Ref.~\cite{HummerSzabo1}), which therefore is valid also for general phase space mappings.

In the case of a molecular dynamics simulation, the mapping $\phi(x)$ consists of a concatenation of a certain number of time steps. Equation (\ref{equ:HuSza1}) can be applied at each stage of the mapping and, accordingly, we rewrite it as
\begin{equation}
\label{equ:HuSza2}
e^{-\beta F(q)}=\frac{Q(\lambda_A)}{Q_0} e^{\beta V(q, \lambda_i)}\langle \delta [q-q(\phi_i(y))] e^{-\beta W_{\phi_i}}\rangle.
\end{equation}
Here, the index $i$ refers to the number of steps the mapping $\phi_i$ and $W_{\phi_i}$ consists of is the work accumulated in the first $i$ steps. The parameter $\lambda$ is changed in discrete steps $\lambda_i$ such that $\lambda_0=\lambda_A$ and $\lambda_n=\lambda_B$ and $n$  is the total number of steps.

In principle, the full free energy profile can be obtained from data at one single time. For better accuracy one can combine the free energy profiles for all time steps using the weighted histogram technique introduced by Ferrenberg and Swendsen \cite{Ferrenberg89}:
\begin{equation}
\label{eq:humCor}
e^{-\beta F(q)} = \frac{\sum_i\frac{\langle \delta[q-q(\phi_i(x))]\exp(-\beta W_{\phi_i})\rangle}{\langle \exp(-\beta W_{\phi_i})\rangle}}
{\sum_i \frac{\exp (-\beta V(q; \lambda_i))}{\langle\exp(-\beta W_{\phi_i})}}.
\end{equation}
Combining the histograms for different times in this way, each histogram contributes most where it has the highest accuracy. In conclusion, we have demonstrated that the method of Hummer and Szabo can be transferred essentially without changes to the case of large time step trajectories. 


\section{Model and Pulling Process}
\label{sec:model}

\begin{figure}[hb]
\centering
\includegraphics[clip=true,scale=.6]{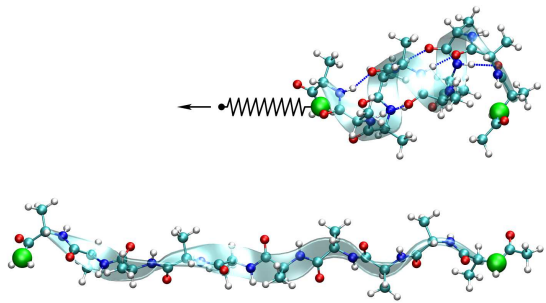}
\caption{(Color online) Deca alanine in its initial $\alpha$-helical (top) and its final coil configuration (bottom). The green atom on the right hand side stays fixed throughout the simulation while the green atom on the left hand side moves in a harmonic trap that is translated from right to left - as indicated by the spring.}
\label{fig:d-ala}
\end{figure}

To investigate the computational efficiency of pulling simulations with large time steps we study the force-induced unfolding of a deca-alanine molecule in vacuum. This oligo-peptide with an acetylated N-terminus and an amidated C-terminus consists of ten alanine residues and forms an $\alpha$-helix that is stable in vacuum at room temperature (Fig.~\ref{fig:d-ala}).  It has $N=109$ atoms and is therefore small enough to permit the calculation of a large number of trajectories in a reasonable amount of time, but is, on the other hand, close to the systems treated in molecular biology or chemistry. Except the differing termini, the model is identical to the one studied by Park \textit{et.~al.}~\cite{PKATS1}. By using this system studied previously as our test case, we can check the results obtained with large time steps not only against our own calculations with ``safe'' time steps, but against an external reference.

The unfolding process from the helix to the coil configuration is induced by imposing a time-dependent trap potential $V(q; t)$ on the amide nitrogen atom of the capped C-terminus at position $q$ while keeping the nitrogen atom of the first residue fixed at the origin. The harmonic trap potential $V(q; t)$ is given by
\begin{equation}
\label{eq:Uguid}
V(q; t) = \frac{k}{2}[q-z(t)]^2,
\end{equation}
where
\begin{equation}
 z(t) = z_i+vt,
\end{equation}
is the trap position and $k$ is the force constant of the potential which is set to $k=14.38$ kcal/mol $\text{\AA{}}^2$ in all simulations. The trap is moved from its initial ($t=0$) position $z_i$ to its final ($t=\tau$) position $z_f$ at constant speed $v=(z_f-z_i)/\tau$. The initial trap position of $z_i=13$ \AA{} is chosen such that the system starts in a slightly compressed state where the end-to-end distance  is smaller than the equilibrium distance, which lies at about $15.4$ \AA{}. The final position $z_f=33$ \AA{} leads to a fully unfolded state. The trap speed $v$ must be chosen sufficiently low such that it allows good sampling using the Jarzynski equality. On the other hand, it should be high enough to permit sampling of sufficiently many trajectories. Following Park \textit{et al.~}\cite{PKATS1}, we chose a trap speed of $0.01$ \AA{}/ps, which amounts to a trajectory length of $2$ ns.
 
The canonically distributed initial conditions, as required by Jarzynski's theorem, are generated from an equilibrium Langevin molecular dynamics simulation at $z(t)=z_i$ with a small time step of $1$ fs. From this ``base trajectory'' we start a non-equilibrium pulling trajectory every one thousand time steps using the velocity Verlet algorithm.
All our simulations were performed using the CHARMM force field version 27 with both CHARMM and the NAMD simulation package, respectively \cite{CHARMMFF1, CHARMMFF2, CHARMM, NAMD}.


\section{Results}
\label{sec:results}
	
In this section we first show that we obtain identical free
energy profiles (well within
statistical error bars), regardless of the time step used. We then discuss an interesting effect we observed for the temperature behaviour in our large timestep simulations. Finally, we address the central point of this paper, the efficiency of the large time step method for the calculation of free energy profiles.
		

\subsubsection{Free energy profiles}

\begin{figure}[ht]
\centering
\includegraphics[clip=true,scale=0.3]{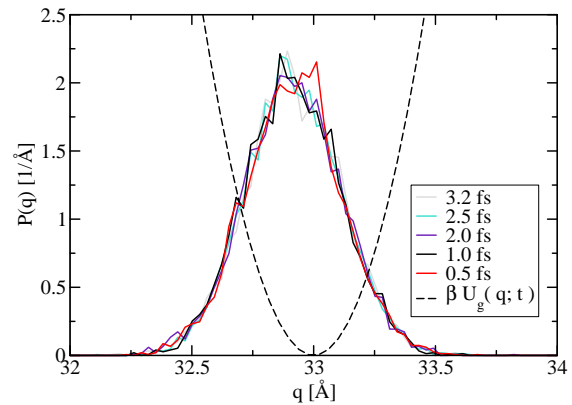}
\caption{(Color online) Histograms $P(q)$ of the end-to-end-distance $q$ for various timesteps obtained from $5000$ trajectories at $t=2$ ns corresponding to the final position of the trap.}
\label{fig:etehist}
\end{figure}

From repeated pulling simulations one can collect separate distributions of the end-to-end distance $q$ for every time $t_i$. Examples of such distributions obtained for the final position of the pulling trap for different time steps $\Delta t$ are depicted in Fig.~\ref{fig:etehist} along with the trap potential. Note that the histograms do not differ significantly for the different time steps and that the distributions are not centred at the minimum of the trap potential but to the left of it, indicating that the molecule slightly lags behind the trap potential.

\begin{figure}[ht]
\centering
\includegraphics[clip=true,scale=0.3]{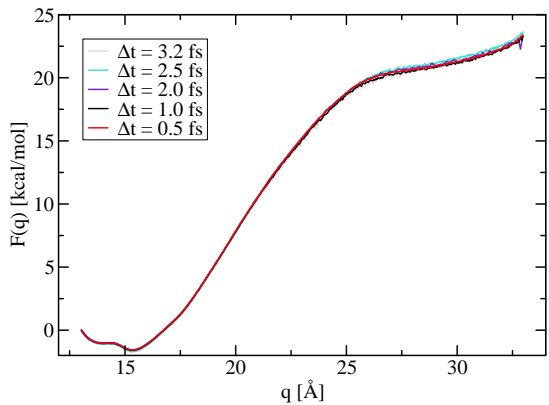}
\caption{(Color online) Free energy profiles $F(q)$ as function of the end-to-end distance $q$ for various timesteps obtained from $5000$ trajectories using the Hummer-Szabo procedure.}
\label{fig:fprof_all_corr}
\end{figure}

Using the procedure of Hummer and Szabo, we have combined the distributions  corresponding to different times obtaining the free energy as a function of the end-to-end distance $q$. For the application of Equ.~(\ref{eq:humCor}) it is important that the histograms of the end-to-end distance have sufficiently narrow bins. If not, the correction 
for the pulling potential in the denominator of Equ.~(\ref{eq:humCor}) varies too much within a bin, thus giving the wrong correction factor. In our simulations we used a  bin size of $0.02$ \AA{}. The free energy profiles calculated with Equ.~(\ref{eq:humCor}) are depicted in Fig.~\ref{fig:fprof_all_corr} and agree well with the curves calculated by Park \textit{et.~al.}~\cite{PKATS1}. As can be inferred from the figure, all time steps yield the same result within statistical deviations. The free energy profile $F(q)$ displays a minimum at an end-to-end distance of $q\approx 15$ \AA{} corresponding to the $\alpha$-helical native state. For increasing $q$, the free energy then grows almost linearly up to $q\approx 25$ \AA{}. In this regime, the hydrogen bonds in the helix are consecutively cleaved. After all hydrogen bonds have been broken, the free energy keeps growing, but at a smaller rate. No stable state exists for the unfolded molecule.

\subsubsection{Artifacts of large time step trajectories}

\begin{figure}[ht]
\centering
\includegraphics[clip=true,scale=0.3]{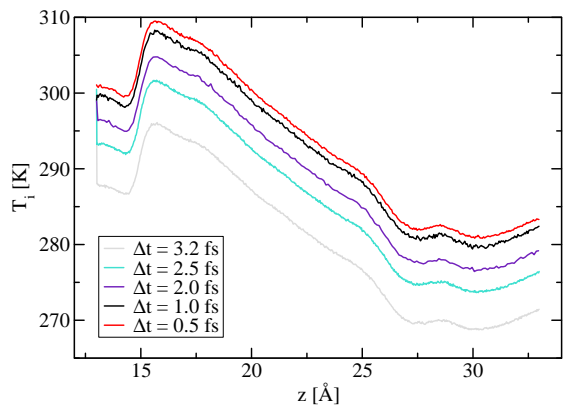}
\caption{(Color online) Instantaneous temperature $T_i$ as a function of the trap position $z$ obtained for various timesteps and averaged over $2000$ trajectories.}
\label{fig:temp-curve}
\end{figure}

The instantaneous temperature
\begin{equation}
T_i= \frac{2 \left\langle E_\text{kin}\right\rangle}{3 N k_\text{B}} ,
\end{equation}
during the pulling process averaged over two-thousand trajectories is depicted in Fig.~\ref{fig:temp-curve} for various time steps. The overall temperature behaviour is similar for all time steps: after a small rise occurring at a trap position of about $15$ \AA{}, the temperature decreases steadily throughout the switching process. However, the temperature traces differ by an essentially constant offset which is due to a very sharp decline of up to five percent from the $300$ K of the Langevin base trajectory. This initial temperature drop, which occurs during the first few integration steps, grows with increasing time step and is accompanied by an increase in the potential energy such that the total energy remains approximately constant. The potential energy increase is mainly caused by an increase of the angle bending potential energy terms and, to a smaller extent, of the bond stretching interactions.

The redistribution of energy from kinetic to potential terms suggests that the equipartition theorem may be violated for large time step trajectories. We indeed find that in straightforward equilibrium MD simulations carried out with time steps larger than $2$ fs the kinetic energy of the hydrogen atoms is up to $\approx 10$ percent less than that of heavier atoms. 

\begin{figure}[ht]
\centering
\includegraphics[clip=true,scale=0.4]{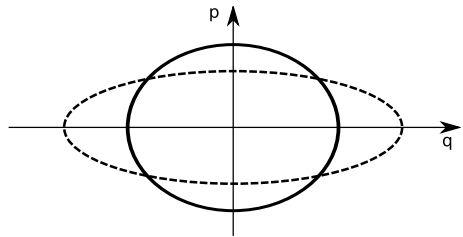}
\caption{Schematic phase space portrait of the exact dynamics of a harmonic oscillator (solid line) and of the dynamics obtained with large time steps (dashed line).}
\label{fig:ellipsoid}
\end{figure}	

This phenomenon is most likely linked to the nearly harmonic potentials used to model the angle bending and bond stretching interactions and can be made plausible by considering a one-dimensional harmonic oscillator with Hamiltonian
\begin{equation}
\label{eq:hoHam}
\mathcal{H}(q, p)=\frac{p^2}{2m}+\frac{m \omega^2 q^2}{2}.
\end{equation}
Here, $q$ and $p$ denote the position and the velocity of the oscillator and $m$ and $\omega$ are its mass and frequency, respectively. As is well known \cite{TOXVAERD1994,FrenkelSmit1,Yoshida93,Saha92}, symplectic integration schemes like the velocity Verlet algorithm do not conserve the actual Hamiltonian of a system, but a timestep-dependent ''shadow Hamiltonian'' $\mathcal{H}_P$ which converges to the real Hamiltonian only in the limit of $\Delta t\rightarrow 0$. For the harmonic oscillator this shadow Hamiltonian is known analytically \cite{GANS2000}:
\begin{equation}
\label{eq:hoSHam}
\mathcal{H}_P(q, p)= \frac{p^2/2m}{1-( \omega \Delta t/2)^2}+\frac{m \omega^2 q^2}{2} .
\end{equation}
Starting from a given phase space point $(q_0, p_0)$ the natural dynamics of the harmonic oscillator traces out an ellipsoid on which the Hamiltonian from Equ.~(\ref{eq:hoHam}) is conserved (solid line in Fig.~\ref{fig:ellipsoid}). However, a trajectory obtained by integrating the equations of motion of the harmonic oscillator with finite time step $\Delta t$ lies on a curve on which the shadow Hamiltonian $\mathcal{H}_P(x, v)$ rather than the Hamiltonian $\mathcal{H}$ is conserved (dashed line in Fig.~\ref{fig:ellipsoid}). It follows from the specific form of the shadow Hamiltonian, that this curve, which is also an ellipsoid, is compressed on the momentum axis by a factor
\begin{equation}
\frac{p_0^2+m^2\omega^2q_0^2 (1-\omega^2\Delta t^2/4)}{p_0^2+m^2\omega^2q_0^2}
\end{equation}
but expanded on the position axis by a factor 
\begin{equation}
\frac{p_0^2(1-\omega^2\Delta t^2/4)^{-1}+m^2\omega^2q_0^2}{p_0^2+m^2\omega^2q_0^2}
\end{equation}
with respect to the ellipsoid defined by a constant Hamiltonian. Thus, increasing the integration time step $\Delta t$ leads to a decrease in the average kinetic energy and, at the same time, to an increase of the average potential energy.

Under the assumption that the dynamics obtained by application of the velocity Verlet algorithm is nearly continuous, a canonical average of the kinetic and potential energy averaged over one oscillation period yields
\begin{equation}
\langle K \rangle = \frac{k_{\rm B}T}{2}\left(1-\frac{\omega^2 \Delta t^2}{8}\right)
\end{equation}
and 
\begin{equation}
\langle V \rangle = \frac{k_{\rm B}T}{2}\left(\frac{1-\omega^2 \Delta t^2/8}{1-\omega^2 \Delta t^2/4}\right),
\end{equation}
respectively. The deviation $\Delta T_{\rm kin} /T$ of the kinetic temperature $T_{\rm kin}=2\langle K\rangle / k_{\rm B}$ from the temperature $T$ of the distribution of the initial conditions is then obtained from the averaged kinetic energy according to:
\begin{equation}
\frac{\Delta T_{\rm kin}}{T}=\frac{T_{\rm kin}-T}{T} = -\frac{\omega^2 \Delta t^2}{8}.
\end{equation}

\begin{figure}[ht]
\centering
\includegraphics[clip=true,scale=0.3]{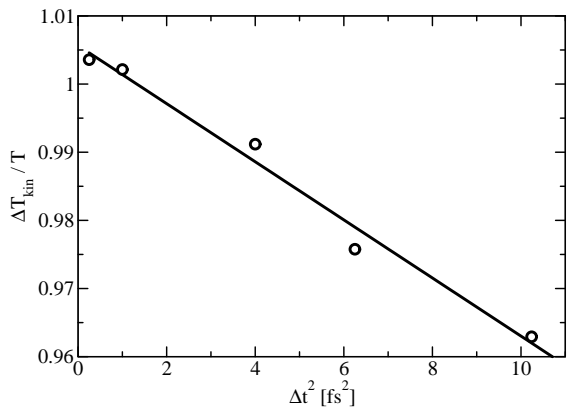}
\caption{Relative deviation $\Delta T/T$ of the kinetic temperature from the temperature of the initial conditions (symbols) as a function of $\Delta t^2$. The line is a linear fit to the data with slope $-0.0043$ fs$^{-2}$.}
\label{fig:trel}
\end{figure}

The quadratic dependence of the temperature drop from the time step expected for a system with nearly harmonic degrees of freedom can be tested for the deca-alanine molecule. Figure \ref{fig:trel} shows the relative temperature drop obtained from the data depicted in Fig.~\ref{fig:temp-curve} as a function of $\Delta t^2$ along with a linear fit to the data. Approximately, the relative temperature drops follow the quadratic behaviour found in the harmonic oscillator. Moreover, the slope of the fitted line is consistent with the assumption that each hydrogen atom is involved in exactly one harmonic degree of freedom and the frequency of the corresponding motion is of the order $\nu\approx 3000 {\rm cm}^{-1}$. 

The above considerations suggest that the sudden drop of the kinetic temperature observed for the deca-alanine with increasing time step may be related to the particular form of the shadow Hamiltonian corresponding to the nearly harmonic degrees of freedom of the molecule. We stress, however, that this effect has no impact on the large timestep method, because the only requirements are a canonical ensemble of initial conditions and a volume preserving integrator such as the velocity Verlet scheme. As long as these conditions are met, the large time step method is valid regardless of the unusual behaviour of some energy components.


\subsubsection{Efficiency}

To assess the efficiency of the large timestep algorithm we need a way to determine the statistical error in the free energy profile as a function of the number of trajectories used in the Jarzynski average. For this purpose we use a block averaging method \cite{AllenTildesley1}, analogous to Zuckerman and Woolf \cite{Zuckerman02}. First we partition all $N$ trajectories into $M$ ''blocks'' of size $n\ll N$ and introduce the free energy $F_n^i(q)$ calculated by applying Equ.~(\ref{eq:humCor}) only to the $n$ trajectories of block $i$. Then, we compute the statistical deviation - Equ.~(\ref{eq:cofN}) - of $F_n^i(q)$ from the mean $\bar{F}(q)$, averaged over all trajectories. To improve the accuracy of this error estimate we average also over the end-to-end distances $q$:
\begin{equation}
\label{eq:cofN}
C^2(n)=\frac{1}{S}\sum_{q}\left\lbrace\frac{1}{M} \sum_{i=1}^{M}\left[ F_n^i(q)-\bar{F}(q)\right]^2\right\rbrace.
\end{equation}
Here, $S$ is the number of end-to-end distances over which the average extends and in all our calculations $S=1000$. The quantity $C^2(n)$ is the mean square deviation of the block free energy from the free energy obtained from all trajectories. For a different approach to efficiency estimation in fast switching simulations, which, however, requires a larger computational effort, see \cite{Lechner1}.

\begin{figure}[ht]
\centering
\includegraphics[clip=true,scale=0.3]{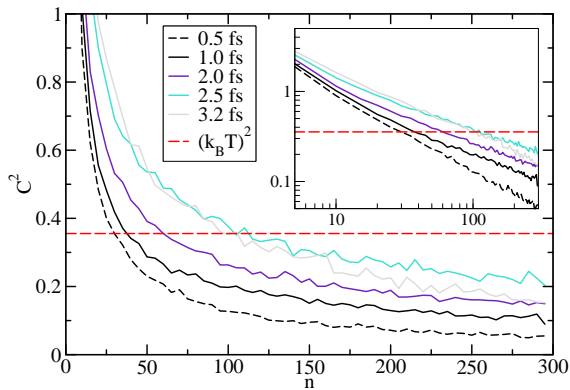}
\caption{(Color online) $C^2(n)$ as a function of the block size $n$ for different timesteps and the threshold $(k_{\rm B}T)^2 \approx 0.355 (\text{kcal/mol})^2$ (red dashed line). The inset shows the same graphs on a doubly logarithmic scale.}
\label{fig:blockavg}
\end{figure}

From the resulting curves $C^2(n)$, which are depicted in Fig.~\ref{fig:blockavg}, we can extract the number $N_{\Delta t}$ of trajectories needed to obtain an accuracy of $k_{\rm B}T$. To do that we determine the intersection of the correlation curves $C^2(n)$ with the error threshold, shown as a red dashed line in Fig.~\ref{fig:blockavg}. The threshold of $k_{\rm B}T$ is arbitrary, but, as the inset of Fig.~\ref{fig:blockavg} indicates, the relative numbers $N_{\Delta t}$ for different timesteps are largely independent of the choice of the threshold.

\begin{figure}[ht]
\centering
\includegraphics[clip=true,scale=0.3]{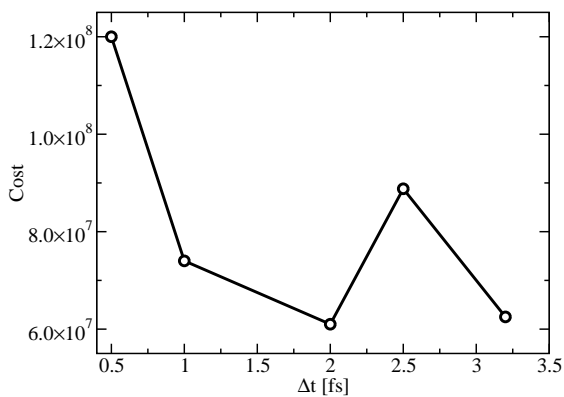}
\caption{Computational cost as a function of time step $\Delta t$.}
\label{fig:cost}
\end{figure}

The total computational cost required to obtain a free energy profile with an accuracy of $k_{\rm B}T$ is proportional to the number of force evaluations performed in the calculation and thus to the number $L$ of time steps per trajectory:
\begin{equation}
\label{eq:cost}
\text{cost}(\Delta t) \equiv L N_{\Delta t} = \frac{\tau}{\Delta t} N_{\Delta t} .
\end{equation}
The cost function $\text{cost}(\Delta t)$ is the total number of molecular dynamics steps required to calculate the free energy with accuracy $k_{\rm B}T$, or, in other words, the total computational cost of the simulation in units of the cost of one single molecular dynamics step. The behaviour of the computational cost as a function of the integration time step $\Delta t$ (see Fig.~\ref{fig:cost}), indicates that in the range $\Delta t = 1.0-3.2$ fs the computational cost of the free energy calculation is essentially constant while time steps shorter than $\approx 1$ fs lead to an increased computational cost.

\section{Concluding remarks}
\label{sec:conclusions}

The generalisation of the Jarzynski theorem for invertible and differentiable maps justifies the use of large time steps in fast switching simulations and thus permits the calculation of valid free energies from computationally inexpensive trajectories. In the present work we have adapted the Hummer-Szabo procedure to large time step dynamics and have shown that it is applicable to the simulation of complex biomolecules.

Approximate large time step trajectories generated with the velocity
Verlet algorithm display an interesting initial temperature drop,
which is most likely due to the specific properties of the shadow
Hamiltonian conserved by the discrete dynamics. Although this effect
leads to inaccurate trajectories violating the equipartition theorem,
it is irrelevant for the validity of the large timestep formalism
since all its requirements, a phase space volume conserving integrator
and a canonical distribution of initial conditions, are met also in
this case. It should be noted that the temperature jump indicating the approximative nature of the trajectories can be observed already at a
time step of 2~fs, a value which is often used in equilibrium simulations of protein
and peptide systems. Thus, one may argue that the large time step
approach is more correct than a comparable calculation based
on equilibrium methods alone.
		
The potential computational benefit of the large time step method is in part compensated by a growth of statistical fluctuations observed in simulations using computationally inexpensive trajectories generated with larger time steps. All time steps from about $1$ fs up to the stability limit of the integrator yield essentially the same efficiency. Time steps smaller than about $1$ fs, however, lead to an efficiency loss.
	

\begin{acknowledgments}
	This work was supported by the Austrian Science Fund (FWF) under grant No. P17178-N02 and within the Science College "Computational Materials Science" under grant W004. Simulations were carried out in part on the Schr\"odinger Cluster of the University of Vienna.
\end{acknowledgments}



\end{document}